\documentclass[12pt]{article}

\input epsf
\def\ba{\begin{eqnarray}}
\def\ea{\end{eqnarray}}
\def\beq{\begin{eqnarray}}
\def\eeq{\end{eqnarray}}
\def\mpl{M_{\rm Pl}}
\def\e{{\epsilon}}

\def\K{{\mathcal K}}

\def\L*{{\cal L}_*}
\def\L{\mathcal{L}}
\def\({\left(}
\def\){\right)}

\def\nn{\nonumber}
\def\p{\partial}
\def\mn{_{\mu \nu}}
\def\lsim{\mathrel{\rlap{\lower3pt\hbox{\hskip0pt$\sim$}}
     \raise1pt\hbox{$<$}}}         %less than or approx. symbol
\def\gsim{\mathrel{\rlap{\lower4pt\hbox{\hskip1pt$\sim$}}
     \raise1pt\hbox{$>$}}}         %greater than or approx. symbol

\usepackage{amsmath}
\usepackage{amsfonts}

\usepackage{graphicx}
\usepackage{bm}        % for math
\usepackage{amssymb}   % for math

\input epsf
\def\beq{\begin{eqnarray}}
\def\eeq{\end{eqnarray}}
\def\e{{\epsilon}}

\def\ba{\begin{eqnarray}}
\def\ea{\end{eqnarray}}
\def\beq{\begin{eqnarray}}
\def\eeq{\end{eqnarray}}

\def\mpl{M_{\rm Pl}}
\def\e{{\epsilon}}

\def\L*{{\cal L}_*}
\def\L{\mathcal{L}}
\def\({\left(}
\def\){\right)}

\def\nn{\nonumber}
\def\z{\zeta}
\def\p{\partial}
\def\mn{_{\mu \nu}}
\def\stu{St\"uckelberg }
\def\mupn{^\mu_{\, \nu}}
\def\<{\langle}
\def\>{\rangle}

\def\lsim{\mathrel{\rlap{\lower3pt\hbox{\hskip0pt$\sim$}}
     \raise1pt\hbox{$<$}}}         %less than or approx. symbol
\def\gsim{\mathrel{\rlap{\lower4pt\hbox{\hskip1pt$\sim$}}
     \raise1pt\hbox{$>$}}}         %greater than or approx. symbol

\def\lsim{\mathrel{\rlap{\lower3pt\hbox{\hskip0pt$\sim$}}
     \raise1pt\hbox{$<$}}}         %less than or approx. symbol
\def\gsim{\mathrel{\rlap{\lower4pt\hbox{\hskip1pt$\sim$}}
     \raise1pt\hbox{$>$}}}         %greater than or approx. symbol

\usepackage{amsmath}
\usepackage{amsfonts}
\usepackage{verbatim}

\begin{document}

\begin{titlepage}

\begin{flushright}
{NYU-TH-07/10/11}

\today
\end{flushright}
\vskip 0.9cm

\centerline{\Large \bf Comments on (super)luminality}
\vskip 0.7cm
\centerline{\large Claudia de Rham$^{a,b}$, Gregory Gabadadze $^c$, and Andrew J. Tolley$^b$}
\vskip 0.3cm

\centerline{\em $^a$D\'epartment de Physique  Th\'eorique and Center for Astroparticle Physics,  }
\centerline{\em Universit\'e de  Gen\`eve, 24 Quai E. Ansermet, CH-1211  Gen\`eve}

\centerline{\em $^b$Department of Physics, Case Western Reserve University, Euclid Ave,
Cleveland, OH, 44106, USA}

\centerline{\em $^c$Center for Cosmology and Particle Physics,
Department of Physics,}
\centerline{\em New York University, New York,
NY, 10003, USA}

\centerline{}
\centerline{}

\vskip 1.9cm

\begin{abstract}

Recently, in an interesting work arXiv:1106.3972  a
solution of the equations of motion of massive gravity was discussed, and it was shown  that one 
of the fluctuations on that solution  is superluminal.
It was also stated that this rules out massive gravity. Here we find  that
the solution itself is rather unphysical. For this we show  that there is another
mode on the same background which grows  and overcomes 
the background in an arbitrarily short period of time, that can be excited 
by a negligible  cost in energy.  This solution is  triggered  by the parameter 
governing the superluminality. 
Furthermore, we also show that the solution,  if  viewed  as a perfect
fluid,  has no rest frame, or that the Lorentz transformation that is needed to boost to
the rest frame is superluminal itself. The stress-tensor of this fluid has
complex eigenvalues, and could not
be obtained from any physically sensible matter.
Moreover, for the same setup we find  another background solution,
fluctuations of which are all stable and subluminal.
Based on these results,  we conclude that the superluminality found
in arXiv:1106.3972  is an artifact of using an inappropriate background,
nevertheless, this solution represents an instructive  example for
understanding  massive gravity. For instance,  on this
background  the Boulware-Deser ghost  is absent, even though this
may naively appear not to be the case.

\end{abstract}

%\vspace{3cm}

\end{titlepage}

\section{Introduction and summary}

Massive gravity (classical theory) has had a turbulent past and present.  To briefly account for
works immediately relevant to the present paper:  Fierz and Pauli (FP)
constructed a ghost-less  and tachyon-free  linear theory \cite {FP}. Van Dam  and Veltman,
and Zakharov (vDVZ),  have independently shown  \cite {vDVZ}  that the FP theory
has discontinuity
in the zero mass limit, and argued that this  excludes massive gravity. Soon after,
Vainshtein showed that the vDVZ discontinuity is an artifact of the  perturbative expansion that
breaks precociously, and argued that upon inclusion of nonlinear terms there should be
nonperturbative continuity to the massless theory, at least for the physical
systems of observational relevance,  thus evading  the vDVZ conclusion \cite {Arkady}.
However, subsequently Boulware and Deser (BD) \cite {BD}
showed that in a broad class of nonlinear extensions  of the FP theory
one is not able to retain the needed five degrees of freedom of a massive graviton;
instead,  the sixth mode becomes propagating  on certain  backgrounds. This mode
typically has negative energies,  and is referred as the BD ghost.

More modern developments were triggered by the DGP model \cite {DGP}, for which it  was
argued by Deffayet et.al.  \cite {DDGV} that the Vainshtein recovery does take place
for sources of
observational interest. This was followed by  a covariant
effective field theory formulation of massive gravity by Arkani-Hamed, Georgi, and
Schwartz \cite {AGS}, who also proposed a  program  to construct a theory that would
avoid the sixth mode  (the BD ghost), starting  from  the analysis of the decoupling limit
\cite {AGS,Ratt,Nicolis,DeffayetRombouts},  where things are easier to handle.

A positive progress toward this goal was made only recently:  in Ref.~\cite {CdRGG}
it was shown  that  in the decoupling limit the BD ghost can be avoided order-by-order
to all orders. The absence of the BD ghost in the decoupling limit is a necessary consistency
condition, but also  turned   out to be a powerful
requirement leading  to  resummation  of an infinite number of terms of the effective
theory, resulting in a covariant Lagrangian  with just a few terms \cite {dRGT}.
The obtained Lagrangian reads:
\beq
\mathcal{L}=\frac{\mpl^2}{2} \sqrt{-g} \left ( R + {m^2} (
\mathcal{L}^{(2)}_{\rm der}(\K) + \alpha_{3} \mathcal{L}^{(3)}_{\rm der}(\K) +
\alpha_{4}  \mathcal{L}^{(4)}_{\rm der}(\K) )  \right )\,.
\label{Lmg}
\eeq
The tensor $\K$ is defined as follows
\beq
\label{Kmn}
\K^\mu_\nu \,=\,
\delta^\mu_\nu -\sqrt{g^{\mu\alpha}\partial_\alpha \phi^{a} \partial_\nu \phi^{b} \eta_{ab}}\,,
\eeq
where the square root above denotes a matrix element of
the root of the matrix; $\eta_{ab}={\rm diag}(-1,1,1,1)$, and $\phi^a(x),~~a=0,1,2,3$ are four spurious
\stu scalar fields  introduced as a redundancy to provide for manifestly covariant
description of massive gravity (for earlier  works  introducing these scalars, see,
\cite {Siegel}.)  Finally, the  mass and potential terms in (\ref {Lmg}) read as follows:
\beq
\label{L2der0}
\mathcal{L}^{(2)}_{\rm der}(\K) &=&[\K^2]-[\K]^2\,,\\
\label{L3der}
\mathcal{L}^{(3)}_{\rm der}(\K)&=&[\K]^3-3 [\K][\K^2]+2[\K^3]\,,\\
\label{L4der}
\mathcal{L}^{(4)}_{\rm der}(\K)&=&[\K]^4-6[\K^2][\K]^2+8[\K^3]
[\K]+3[\K^2]^2-6[\K^4]\,,
\eeq
where we use the notations  $[\K] \equiv ({\rm Tr}\, \K^\mu_\nu )$,  $[\K]^2 \equiv ({\rm Tr}\, 
\K^\mu_\nu )^2$,
while  $ [\K^2 ] \equiv {\rm Tr}\, (\K^\mu_\nu \K^\nu_\alpha)$. The terms $\mathcal{L}^{(n)}_{\rm der}$
give total derivatives upon substitution $\K^\mu_\nu \to \p^\mu\p_\nu \pi$,  as indicated in their notation.

The above  Lagrangian (\ref {Lmg}) has three free parameters (one of them being the graviton mass $m$),
and  for some values of these parameters the theory has been shown to be  free of the BD ghost
away from the decoupling limit up to (and including) the quartic order in
nonlinearities \cite {dRGT}.  Remarkably, Hassan and Rosen \cite {HR}
have managed recently to show that it is  free of the BD ghost away from the decoupling limit,
to all orders\footnote{Our results \cite {dRGT}, as well as
the results of  \cite {HR},  are in conflict with the claim  of Ref.~\cite {Mukhanov}
of the existence  of the BD ghost in the quartic order.  This controversy is addressed
in  Ref.  \cite {dRGT_res2}, where it is shown that \cite {Mukhanov} missed a constraint.
See also discussions  in  Section 2.}.
We note  that the absence of the BD ghost guarantees the absence of the sixth mode.
This however, does not prohibit  one or more of the physical 5 polarizations
to flip the sign of their  kinetic terms on certain backgrounds  and become ghosts.
Such backgrounds should be considered unstable in the theory, but such cases should be
distinguished from the ones  with the sixth mode. For some  work on cosmology
and spherically symmetric solutions in massive gravity see, e.g.,
\cite{dRGHP} - \cite{dRH},
and  Ref. \cite {KurtRev} for a theory review.

\vspace{0.1in}

Recently, in a brief  work Gruzinov \cite {AG} has  found a certain
solution of the theory  (\ref {Lmg}),  and showed that there is a
fluctuation about this solution  which  is  superluminal. Based on this observation,
it was concluded that massive gravity is ruled out.
Below we examine this conclusion more carefully. In Section 2
we show that there exist a growing solution  which overcomes the background arbitrarily 
quickly, and can be excited with virtually no cost in energy.
Moreover, this solution is not related to the BD ghost, as the latter is absent
in this theory.  In Section 3 we show that the solution of \cite {AG}, 
if interpreted as a perfect fluid, has a stress-tensor with complex eigenvalues. Hence, the
rest frame for this fluid can only be  achieved via
superluminal boosts.  As such,  this configuration could not,
as an exact solution, be  obtained from any known physically
meaningful form of matter. In the appendix we discuss the decoupling limit of the
linearized fluctuations.

The present work does not claim to exclude all possible superluminalities in massive gravity.
Indeed, some of the  terms obtained  in the decoupling limit of massive gravity
resemble the Galileon theories  \cite {galileon},   which were shown
to exhibit superluminalities.  It is therefore reasonable to expect that
when massive gravity reduces to a Galileon theory in the decoupling limit,
the fluctuations of the helicity-0 mode around a spherically symmetric solution
can also be superluminal.
However,  for more generic values of the parameters of the theory (\ref {Lmg}), in
particular when $\alpha_3+4\alpha_4\ne0$,  the decoupling limit of the theory cannot
be written explicitly in a Galileon form, and it is possible that the decoupling
limit does not capture the entire physics of the system. In that case a more careful
treatment is required, and it is yet unclear whether the superluminalities around
non-trivial backgrounds survive. We plan to report on this issue in a future work.

% \newpage

\section{Superluminality and growing solutions}

The system of equations of the theory (\ref {Lmg}) reads as follows:
\beq
G_{\mu\nu}(g) + m^2 X_{\mu\nu}(g, \phi) =0\,, \label{eineq} \\
m^2\,\nabla^\mu X_{\mu\nu}(g, \phi)=0\,.  \label{scalareq}
\eeq
Here, $X\mn$ is a tensor  obtained by variation of
the  mass terms in (\ref {Lmg}), and is given explicitly in Section 3. 
Ref.  \cite {AG}  considers a classical scalar field configuration
\beq
\phi_{cl}^a =(\phi^0, \phi^1, \phi^2,   \phi^3)_{cl} =  (t, x+\e  t, y,z),
\label{phi}
\eeq
with an arbitrary constant $\e$,  and studies a  flat space fluctuation  of
$\phi^2 $ in the $x$ direction,  showing  that this
fluctuation  is  superluminal. Based on this,  the work states
that massive gravity is ruled out.

\vspace{0.1in}

We start by presenting the results  of Ref. \cite {AG} in more detail.
We do this for a small value of $\epsilon\ll 1$, which is enough for
our purposes. For this consider the field configuration  (\ref {phi}).
It produces some stress-tensor
$X\mn$ which also depends on the metric $g\mn$;  since  $X\mn$ is  multiplied
by $m^2$ in eq. (\ref {eineq}),  one assumes that the back-reaction
of $m^2X\mn$  on the metric is negligible.
In this approximation,  the remaining equation is just
an empty space  Einstein equation, which certainly has a solution $g\mn =\eta\mn$.
Hence,  to summarize  the solution of \cite {AG}:
\beq
g_{\mu\nu} =\eta_{\mu\nu} + {\tilde  h}_{\mu\nu}, ~~~
{\tilde  h}_{\mu\nu}\sim {\cal O}(\e m^2 x^. x^.)\,,
\label{gruzsol}
\eeq
where for convenience we have included  the correction  ${\tilde  h}$, where
$x^.$ denotes some components of $x^\mu$ (these corrections can straightforwardly be
calculated for small $\e$, and take the form,  ${\tilde h}_{01}\propto \e m^2 t^2$,
${\tilde h}_{12}\propto \e m^2 ty$).  Since $m$ can be arbitrarily small,
one can neglect ${\tilde  h}$ in (\ref {gruzsol}),  at least in some region of space and
time,  and consider fluctuations on the approximate background $g\mn \simeq \eta_{\mu\nu}$.

To present the results of \cite {AG} more explicitly, we write down the quadratic
Lagrangian  for the fluctuations $\zeta^a$  of  the  four components
of the $\phi^a$ field, while freezing all the other fields in the theory
(the approximation in which this is justified will be discussed later,
see below and the appendix):
\beq
\phi^a = x^a + \delta^a_1\e t  + {\zeta^a(t,x,y,z)\over \mpl  m}.
\label{fluctuations}
\eeq
The Lagrangian for $ \zeta^a$  follows from an expansion of
(\ref {Lmg}) on the background (\ref {phi}, \ref {gruzsol}), and in the quadratic
approximation for the fluctuations $ \zeta^a$
takes the form\footnote{Note that $ \zeta^a$, in spite of
its appearance, does not transform as a vector under diffeomorphisms, instead, it
transforms as a four-coordinate.}:
\beq
\mathcal{L}_{\zeta}=-{1\over 4} F_{\mu\nu}^2(\zeta) -  {3\beta \e \over 2}
F_{1\alpha}F^{\ \alpha}_0 -   \frac \e2  F_{01} (\p_\alpha \z^\alpha)+{\cal O}(\e^2)\,,
\label{Lzeta}
\eeq
where $\beta=-(\alpha_3+1/6)$, and  the total derivative terms have been ignored. Furthermore,
Ref. \cite {AG} focused on the $\z_2$ component propagating only in the $x$ direction
(this does not excite
other fluctuations in (\ref {Lzeta})). For $\zeta_2(t,x)$ the above Lagrangian reduces to
\beq
\mathcal{L}_{\zeta_2}={1\over 2} {\dot \z}^2_2 -{1\over 2} {\z^\prime }^2_2 -  {3\beta \e \over 2}
{\dot \z  }_2 {\z^\prime }_2 +{\cal O}(\e^2)\,,
\label{z2}
\eeq
where an over-dot and prime  denote $t$  and $x$ derivatives respectively.
The dispersion relation that follows to leading order in $\e$,
$\omega \simeq  p ( 1- 3\beta\e /2)$, is superluminal, since
for any nonzero $\beta$, the value of $\e$ can always be chosen to give
superluminality \cite {AG}.

If the Lagrangian (\ref {z2}) is taken in isolation of all the other fields
and interactions, as done in \cite {AG},   then, the derived superluminality can be removed by a
simple change of coordinates to ${\tilde x}  $ and ${\tilde t} $ where,
${\tilde x} =  x+t (3\beta \e/2)$  and ${\tilde t}= t$ (which is just a galilean transformation
with velocity equal to $-3\beta \e/2$). Likewise, from  an innocent field theory
of a scalar $\varphi$ coupled to a source $J$ with the Lagrangian,
$-(\partial_\mu\varphi)^2 +\varphi J$,  one can get the Lagrangian
of the type (\ref {z2})  by the above
change of coordinates.  Hence, if  (\ref {z2}) were the entire Lagrangian
one could quantize fluctuations in the $\{{\tilde x}, {\tilde t}\}$
coordinate system  where no superluminality would appear.

The actual question, however,  is what and how these fluctuations couple to other
fluctuations and external sources, and what those other fluctuations do. In the full theory
the field $\z$ does mix with the tensor and scalar modes at the linearized level, and has also
nonlinear interactions.  Also, there are ${\cal O}(\e^2)$ terms neglected in
(\ref {z2})\footnote{We thank Mehrdad  Mirbabayi who pointed out to us that the Lagrangian
(\ref {Lzeta}), if considered in isolation, has a gauge symmetry, since
the last term in it can simply be removed to the next,  $\e^2$ order,  by a field redefinition,
$\z_0 \to \z_0 -{\e\over 2} \z_1$, $\z_1 \to \z_1 -{\e\over 2} \z_0$. This symmetry,
is not present in the $\e^2$ order, and also in the full massive theory,
due to the coupling  of $\zeta_a$ to the tensor mode.}. We study in turn all
the  fluctuations omitted in (\ref {z2}).

\vspace{0.1in}

Let us first focus on other components of $\z^a$ which were not
considered  in  \cite {AG}. Dropping the last term in (\ref {Lzeta}),  we write
the equations of motion  in the Lorentz gauge $\p_\mu \z^\mu =0$:
\beq
\square (\z_0 - q \z_1) + q \p_0\p_1 \z_0 - q \p^2_0\z_1 =0\,, \nn \\
\square (\z_1 + q \z_0) + q \p_0\p_1 \z_1 -q \p^2_1\z_0 =0\,,  \nn \\
\square \z_{b} + q ( -\p_1\p_b \z_0 +2 \p_1\p_0 \z_b -\p_0 \p_b\z_1) =0\,,
\label{zeqs}
\eeq
where $b=2,3,$ and  $q\equiv 3\beta\epsilon/2$. These empty-space equations have many growing
solutions. The one we focus on is
\beq
\zeta_1  \simeq  {1\over 2} q m_0^3\,  t^2 + m_0^3 t(x-x_0) +
{\cal O} \( \e^2 \),~~~~~ \zeta_2  \simeq  -    m_0^3\, t(y-y_0)  +
{\cal O} \( \e^2 \) \,,
\label{sol01}
\eeq
where $m_0,x_0,y_0,$ are  arbitrary integration constants, and other components of
$\zeta$ are set to zero. In the leading order in
$\e$,  the $\e^2 $ pieces  in the above expressions should be ignored.
Note that for (\ref {sol01}),  $F_{01}= qm_0^3 t + m_0^3 (x-x_0)$, and
$F_{02}=-m_0^3 (y-y_0)$. At each point in space one could choose corresponding 
$x_0,y_0$, such that $F_{01}= qm_0^3 t$ and $F_{02}=0$  point-by-point in the whole space. 

There are some important comments to be made about the solution.

(a) The solution (\ref {sol01}) grows in time $t$ and overcomes the background
(\ref {phi}) for $t \gsim t_*$, where $t_* \equiv (m \mpl/ m_0^3)$. However,
the mere existence of this solution  cannot be interpreted as an instability of the
background (\ref {phi}). The reason is that the solution (\ref {sol01}) has a
nonzero energy density proportional to $F^2_{01} + F^2_{02}$, and in order to excite this
field configuration within a finite volume in space one would need some energy; if such an
energy is supplied, it is then not surprising that (\ref {sol01}) does overcome the background 
after some  period of time.

Nevertheless, there is an aspect of the  solution (\ref {sol01}) that suggests that
the background (\ref {phi}) is unphysical.  To see this let us introduce a
length scale $L_* \equiv q t_*$.  Let us now consider  a small imaginary box
of volume $L^3$ centered around the  point $x=x_0,~~y=y_0,~~z=0$.  Most importantly,
we take $L\lsim L_*$.  Furthermore,  imagine  that we supplied enough energy
in the box,   and the appropriate boundary conditions at its sides,
so that inside the box the solution (\ref {sol01}) is excited, while outside of the box
the background is still given by  (\ref {phi}). Let us now estimate  how  much
energy density we need to supply for this to be the case. The average energy density
in the box will consist of three terms,  $D = D_1+D_2+D_3$, where $D_1 \sim q^2 m_0^6 t^2$,
$D_2 \sim  q t m_0^6 L$, and $D_3 \sim   m_0^6 L^2$. Now, for any time moment
$t> L/q$, the $D_1$ term dominates. What is important, however, is
that $L/q  \lsim L_*/q =t_*$. Hence, at the time moment $t\sim t_*$,
when the solution (\ref {sol01}) in the box
begins to dominate over the background (\ref {phi}), the energy density
that is required to excite it is of the order $D_1\sim q^2$; however, the latter happens to be
zero in our approximation since we are ignoring  terms of order $\e^2\sim q^2$ in the action.
Therefore, we conclude that at the expense of the  energy density that
is zero in our approximation,
we can excite the solution (\ref {sol01}) in a volume of size $L^3 \lsim L_*^3$,
and this solution overcomes the background (\ref {phi})  after $t\sim t_*$ time.
Since $x_0$ and $y_0$  in (\ref {sol01}) are arbitrary, we can now consider the
whole space populated by  non-overlapping boxes in each of which 
an appropriate value of the parameters $x_0,y_0$  are chosen, and 
as a result,  the growth  described above develops in each 
of these boxes. Then, it is logical  to interpret  $t_*$ as a characteristic  
time for the growing solution  to dominate in the entire space.  
Most importantly, this time scale is independent of $q$, and can be made   arbitrarily small
by adjusting the integration constant $m_0$.  Moreover,  the amount of the energy density
needed to excite this solution,  $D_1\sim q^2 m_0^4 t^2$, although negligible in our approximation, 
in any event is much smaller than the characteristic scale of  the stress-tensor for the 
background,  $\e m^2\mpl^2 $, as long as $t< (m\mpl/(\e m_0^3))$; the latter is always the case,  
for $t\lsim t_*$. \\

(b) Although, the above growth can be arbitrarily fast, it is triggered  by the
same parameter $q\sim \e$ that sets the superluminality of the $\zeta_2$ mode. 
This phenomenon  disappears in the limit $q\to 0$, even though
for $q=0$ the solution  (\ref {sol01}) still grows and it may appear that
the above-given argument would  still hold in the $q\to 0$ limit.
In this limit the solution (\ref {sol01}) reduces
to the terms  $m_0^3 t(x-x_0)$ and $m_0^3t(y-y_0)$, but no $t^2$ term is remaining.
Then, within a box  of size $L^3$ (which is now necessarily larger than $L^3_*$ that
tends to zero)  one would be able to excite the solution at the cost of a
non-negligible energy density  $D\sim D_3\sim m_0^6 L^2$.
This energy density cannot be ignored  as there are no neglected
$q^2$ terms in the action any more. Clearly, such a solution does not indicate
any problem  of the background, it simply reflects the fact that the background has
changed at the expense of the supplied finite energy density \footnote{This is similar to the
case of a free massless scalar field which has a solution, $\phi \sim t$;
in any infinitesimal region of space one needs to supply a  nonzero energy density to
excite this solution. Likewise, to excite the solutions, $\phi \sim tx,ty$, which
also exist in this theory, one would need finite energy density in any finite volume.}.\\

(c) The Lagrangian (\ref {Lzeta}), as was pointed out above,  is not
the total  Lagrangian of our theory,  even at the linearized level and even
in the leading order in $\e$. Then the question  arises of whether the mixing with other fields
plays any role. Since this is a bit technical,
we address this question in the appendix, where we show that only
in the $m\to 0$ limit the dynamics of
the $\zeta_a$ modes in  (\ref {Lzeta}) can be decoupled
from the tensor  and scalar modes, assuming that one
ignores nonlinear interactions as well. \\

In spite of the above described issue with the background  (\ref {gruzsol})
it is important to emphasize that this problem is not related in any
way to the Boulware-Deser mode, i.e., to  a potential sixth degree of freedom
in a broad class of  massive gravities, which is absent in the present
model (\ref {Lmg}). To see that the sixth
mode is not propagating on the background considered in \cite {AG},
we look at the  full action for the tensor fields.
In unitary gauge $\phi^a = x^a$, the solution considered in \cite{AG} amounts to taking
a background
solution for the metric which is Minkowski, but not in Cartesian form. Specifically the
background
metric in unitary gauge is $ds^2 = -dt^2 + (dx-\epsilon dt)^2 + dy^2 + dz^2$. Expanding
to quadratic
order in perturbations around this solution and to first order in $\e$, the mass term which 
is now expressed entirely in terms of the tensor field since $\zeta^a=0$ is
\beq
-{m^2\over 8} \left ( {\bar h}_{\mu\nu}^2 -{\bar h}^2 +6c_1
\e {\bar h}_{1\alpha}{\bar h}^\alpha_0 - (6c_1+1)\e
{\bar h}_{01} {\bar h}  \right )\,,
\label{barh}
\eeq
with $c_1=-\alpha_3-3/2$.
We can use this form in order to count the physical degrees of freedom. 
In order for the BD ghost to be absent, one has to have
the Hamiltonian constraint \cite {BD}. In the FP linearized theory the Hamiltonian constraint
is enforced by $h_{00}$ being a  Lagrange multiplier, while $h_{0j}$ is  algebraically determined by an
equation that is {\rm independent} of $h_{00}$. In the Lagrangian (\ref {barh}), however,  $h_{00}$
mixes with $h_{01}$,  and this may seem to forbid the presence of a constraint. However, this is not so,
there still exists a linear combination of the fields that is a Lagrange multiplier in the approximation used.
A convenient way to see this is to calculate the determinant of the $4\times 4$ Hessian matrix  for the Lagrangian
${\cal H}^{\mu \nu}\equiv {\delta^2{\cal  L}\over \delta h_{0\mu} \delta h_{0\nu}}$.
If the determinant is zero, then  there are constraints.
It is straightforward to calculate  that the determinant of the Hessian that follows from (\ref {barh})
is  of order $\e^2$, i.e., it is zero in  our approximation, while the rank of the Hessian is 3.
Hence, there is one constraint in the system. Moreover, conservation of this constraint
leads to a secondary constraint, as shown  for these theories exactly in \cite {HR};
due to these one is able to eliminate the BD ghost.

One may also consider the counting of degrees of freedom in the non-unitary gauge considered in \cite{AG}.
In such a gauge the tensor mode propagates two degrees of freedom, while there should be
only three degrees of freedom for the four \stu fields. The latter requirement  at first sight seems
unlikely  since there is no gauge invariance for $\zeta$ in  the full theory,
and moreover, $\zeta_0$ enters with a time derivative, even in the simplest  case of $c_1=0$.
Based on this one may be tempted to conclude that the Lagrangian  for the
$\zeta$ field (after gauge fixing $h\mn$)
propagates 4 degrees of freedom. However, by
more careful inspection one can  show that   there are constraints
that render  only 3 degrees of
freedom in the $\zeta$ sector in general
(detailed discussions of how this works in the full nonlinear theory
are given in \cite {dRGT_res2}).

\vspace{0.1in}

In conclusion, the solution (\ref {gruzsol}) seems  problematic, despite being ghost-free.
Is there another problem-free solution for the very same  configuration
of the \stu fields (\ref {phi})?  The answer is positive.  It is
straightforward  to find  another solution to the system of eqs. (\ref {eineq})
and (\ref {scalareq}),   for  given  (\ref {phi}):
\beq
g_{\mu\nu} =\eta_{\mu\nu} +\e (\delta_\mu^0 \delta_\nu^1 +
\delta_\mu^1 \delta_\nu^0)+\mathcal{O}(\e^2)\,.
\label{exact}
\eeq
or exact to all orders $d s^2=-d t^2 + (d x + \epsilon\ d t)^2+d y^2+d z^2$.
The above solution differs from  (\ref {gruzsol}),
by  $\e$, i.e., by the same parameter that sets
superluminality found in \cite {AG}.
Furthermore, it is easy to notice that the solution (\ref {phi},\ref {exact})
is nothing but  the Minkowski solution,  $ g_{\mu\nu}= \eta_{\mu\nu},~~\phi^a =(t,x,y,x)$,
transformed by the coordinate change $x^\mu \to x^\mu + \e t \delta^\mu_1$.
Therefore, the fluctuations above the solution (\ref {phi},\ref {exact})
are just  ordinary fluctuations of the Fierz-Pauli theory,
\beq
-{m^2 \over 8} \eta^{\mu\alpha}  \eta^{\nu\beta} \left (h_{\mu\nu} h_{\alpha \beta } -
h_{\mu\alpha}  h_{\nu\beta} \right)\,,
\label{FP}
\eeq
which  are known to be subluminal and stable.

\section{Superluminality of the source}

The solution considered in \cite {AG} is not an exact solution of massive gravity.
As we have explained, it is at best a solution valid locally in a
space-time region whose size/time scale
is set by $L \sim 1/({\sqrt{\epsilon}m})$. Alternatively we can allow it to be an exact
solution by adding an external source $T^{\rm ext}_{\mu\nu}$
which is chosen so that
\beq
m^2 X_{\mu\nu}=T^{\rm ext}_{\mu\nu}.
\eeq
In principle we could imagine this external source being set up by a configuration of matter,
a fluid, or a set of scalar or gauge fields. However it is easy to see that the `fluid' needed would itself be composed of
superluminal matter. To see this, imagine $T^{\rm ext}_{\mu\nu}$ were described by a perfect fluid. Let
us assume that the fluid has a rest frame. If this is the case we can perform a Lorentz
transformation so that in the vicinity of one point the fluid has zero velocity. At that point
$T_{\rm ext}^{0i}=0$. Since the background metric is flat, $g_{0i}=0$ and so $T_{{\rm ext} \, i}^0=T_{{\rm ext} \, 0}^i=0$.
This in turn implies that the energy density $T^0{}_{0\, {\rm ext}}$ is one of the eigenvalues of the stress energy tensor
$T^{\mu}{}_{\nu \, {\rm ext}}$. The stress-energy tensor is expressed in terms of the tensor $\mathcal{K}\mupn$ in the combination
\ba
&&X\mn=\K g\mn -\K\mn+(1+3\alpha_3)\(\K\mn^2-\K \K\mn+\frac 12 \([\K]^2-[\K^2]\)g\mn\)\\
&&+\alpha \(\K\mn^3-\K \K\mn^2+\frac 12 \K\mn \([\K]^2-[\K^2]\)-\frac 16 ([\K]^3-3[\K][\K^2]+2[\K^3])g\mn\)\,,\nn
\ea
where $\alpha \equiv \alpha_3+4 \alpha_4 $. So the eigenvalues of
$T^{\mu}{}_{\nu \, {\rm ext}}$ are determined by the eigenvalues
$\lambda_K$ of $\K\mupn$, which in turn are expressed as
\ba
\lambda^{(n)}_{\K}=1-\sqrt{1-\lambda^{(n)}_Y} \hspace{20pt}{\rm for }\ n=1,\ldots,4\,,
\ea
where $\lambda_Y$ are the eigenvalues of $Y^{\mu}{}_{\nu}$:
\beq
Y^{\mu}{}_{\nu}=g^{\mu\alpha} \partial_{\alpha} \phi ^a \partial_{\nu} \phi^b\eta_{ab}.
\label{matrix}
\eeq
It is straightforward to show that the first two eigenvalues of this tensor
are complex for the background
solution,  and so $T^0{}_{0 \, {\rm ext}}$ is complex in this frame.
Explicitly for the background considered the  matrix (\ref {matrix})  is
\beq
\left(\begin{array}{cccc}
1-\epsilon^2 & -\epsilon & 0 & 0 \\
\epsilon & 1 & 0 & 0 \\
0 & 0 &1 & 0 \\
0 &0&0&1
\end{array} \right)
\eeq
and its eigenvalues are easily shown to be
\beq
\(\lambda_Y^{(1)},\lambda_Y^{(2)},\lambda_Y^{(3)},\lambda_Y^{(4)}\)=
\(1-\frac{1}{2}\epsilon^2+\frac{i}{2}\epsilon \sqrt{4-\epsilon^2},1-\frac{1}{2}
\epsilon^2-\frac{i}{2}\epsilon \sqrt{4-\epsilon^2},1,1\).
\eeq
We can therefore immediately infer that the eigenvalues of $\K$ are also complex, and so are
the eigenvalues of $T^{\mu}{}_{\nu \, {\rm ext}}$.

This implies that there is no rest frame for the fluid, or that the Lorentz
transformation needed to boost to the rest frame is superluminal (and hence a complex
transformation), since it is not possible to perform a real Lorentz transformation to
set $T^{\rm ext}_{0i}=0$. As such this configuration could not, as an exact solution, be
obtained from any known physically sensible form of matter.

Even in the absence of a source, the same arguments hold. It is clear that for any solution
of the equations $G_{\mu\nu} + m^2 X_{\mu\nu}=0$ which looks locally like flat space-time with
the field profile described in (\ref {phi},\ref {gruzsol}), it is not possible to boost to
a frame in which $G^{0i}=0$ in the local vicinity of a point. In this sense the solution
already at the level of the background looks superluminal and rather unphysical.

\vspace{0.1in}

We would  like to thank L. Berezhiani, G. Chkareuli, S. Dubovsky,
D. Pirtskhalava and R. A. Rosen for useful discussions and comments, and
especially M. Mirbabayi for his valuable input. CdR is supported
by the Swiss NSF and GG is supported by the NSF grant PHY-0758032. GG regrets that
discussions on the subject of \cite {AG} could not be contained within NYU. \\

\subsection*{Appendix}

Below we show that the Lagrangian (\ref {Lzeta}) can be obtained  from the full theory
in the limit $m\to 0$. For this we start with the mass terms on the background
(\ref {phi}, \ref {gruzsol})
\beq
{\cal L}_m = - {m^2 \mpl^2\over 8} \(h_{\mu\nu}^2 - h^2 + \e_1 h_{1\mu } h^\mu_0 -\e_2 h_{01}h+
4\e h_{01} \)\,,
\label{Lm}
\eeq
where $\e_1 \equiv \e(6\beta-4)$ and   $\e_2 \equiv \e(6\beta-3)$, and
all the indices are contracted by $\eta \mn$. We express the Lagrangian
in terms of the \stu fields by using the substitution
\beq
h_{\mu\nu} \to {h_{\mu\nu}\over \mpl} - {S\mn \over m \mpl} -
{\p_\mu \z^a \p_\nu \z^b \eta_{ab}\over  m^2 \mpl^2}\,,
 \label{stu}
\eeq
where we defined $S_{\mu\nu} \equiv \partial_\mu \z_\nu + \p_\nu  \z_\mu +
\e(\delta^0_\mu \p_\nu \z_1 +  \delta^0_\nu \p_\mu \z_1)$, and
introduced  canonical normalizations for all fields.

Then, the total Lagrangian  reads as follows:
\beq
{\cal L} & =&  {\cal L}_{EH}(h) + {\cal L}_\z(\z) + {m\over 4} S^{\mu\nu}
(h\mn -\eta \mn h) \nn \\
& +& {m \e_1 \over 8} \(h_{1\mu} S^\mu_0 + S_{1\mu}h^\mu_0 \) -  {m \e_2 \over 8}
\(h_{01} S + S_{01}h \) + {\cal O} (m^2)\,,
\label{Lexp}
\eeq
where   ${\cal L}_{EH}(h)$ denotes the linearized Einstein-Hilbert term,
while ${\cal L}_\z(\z)$ is the Lagrangian given in (\ref {Lzeta}).
We also ignored the terms  of order ${\cal O} (m^2)$ or smaller, ${\cal O} (\e m^2)$,
in (\ref {Lexp}). Note that the tadpole appearing in
(\ref {Lm}) gets canceled in (\ref {Lexp})  by the corresponding tadpole
coming from the EH term taken on the background (\ref {gruzsol}).

We see that the  fields $\zeta^a $ mix  to the tensor field. Our goal is  to show
that this mixing disappears  in the $m\to 0$ limit.  For this we note that
we   can remove the third term in ${\cal L}_\z(\z)$
by a linear field redefinition (see footnote 3), and  then introduce the helicity-0 field by
the change of variables $\zeta_\mu \to \zeta_\mu  +\partial_\mu \pi /m $.
As a result we get the following Lagrangian:
\beq
{\cal L} &=& {\cal L}_{EH}(h) + {\cal L}_1(\z) + {1 \over 4 }{\cal P}^{\mu\nu} (\pi)
(h\mn -\eta \mn h)  +    \nn \\
&+ & {\e_1 \over 4 } \(h_{1\mu}   \p^\mu \p_0 \pi  +  \p_1 \p_\mu \pi h^\mu_0 \) - { \e_2 \over 4}
\(h_{01} \square \pi  +   \p_0\p_1 \pi h \) + {\cal O} (m,  \e m,  m^2)\,,
\label{final}
\eeq
where ${\cal L}_1(\z)$  is the Lagrangian (\ref {Lzeta}) less the last
term, and  ${\cal P}\mn (\pi) \equiv 2 \p_\mu\p_\nu\pi + {3\e\over 2}(\delta^0_\mu \p_\nu \p_1 \pi
+ \delta^0_\nu \p_\mu \p_1 \pi )  + {\e\over 2}  ( \delta^1_\mu \p_\nu \p_0 \pi
+ \delta^1_\nu \p_\mu \p_0 \pi)$. As we see, there is a mixing between the tensor mode
and the  helicity-0 mode $\pi$. Due to this mixing the helicity-0  gets a kinetic
term via the shift $h_{\mu\nu} \to h_{\mu\nu} +\eta \mn \pi$;
as a result, the helicity-0 would  couple to an external source had we introduced
it in the theory. However,  the    field $\zeta$ does not couple with anybody
in this limit. This would be so even if we were to introduce a  stress-tensor
of an external matter. The coupling of  $\zeta$ appears only at a nonlinear level.

\end{document}